# A high sensitivity HI survey of the sky at $\delta \leq -25°$

## Final data release

E. Bajaja[1]⋆, E.M. Arnal[1,2]⋆, J.J. Larrarte[1]⋆⋆, R. Morras[1,2]⋆, W.G.L. Pöppel[1]⋆, and P.M.W. Kalberla[3]

[1]  Instituto Argentino de Radioastronomía, c.c. 5, (1894) Villa Elisa, Argentina
     e-mail: `bajaja@iar.unlp.edu.ar, arnal@iar.unlp.edu.ar, rimorras@isis.unlp.edu.ar,`
     `wpoeppel@iar.unlp.edu.ar`
[2]  Facultad de Ciencias Astronómicas y Geofísicas, UNLP, Av. Paseo del Bosque S/N, 1900 La Plata, Argentina
[3]  Radioastronomisches Institut der Universität Bonn, Auf dem Hügel 71, 53121 Bonn, Germany
     e-mail: `pkalberla@astro.uni-bonn.de`

**Abstract.** We present the final data release of the high sensitivity $\lambda$ 21-cm neutral hydrogen survey of the sky south of $\delta \leq -25°$. A total of 50980 positions lying on a galactic coordinate grid with points spaced by $(\Delta l, \Delta b) = (0°.5/\cos b, 0°.5)$ were observed with the 30-m dish of the Instituto Argentino de Radioastronomía (IAR). The angular resolution of the survey is HPBW = 0°.5 and the velocity coverage spans the interval -450 km s⁻¹ to +400 km s⁻¹ (LSR). The velocity resolution is 1.27 km s⁻¹ and the final rms noise of the entire database is 0.07 K.
The data are corrected for stray radiation and converted to brightness temperatures.

## 1. Introduction

The motivation, observations and reductions (without stray radiation correction) of the Southern Sky HI Survey, made with the 30-m dish of the Instituto Argentino de Radioastronomı́a (IAR), has been thoroughly described by Arnal et al. (2000). We give here only a brief summary of the main parameters that will be useful for the description of the complementary and final reduction of the survey to be described here.

The survey consists of 50980 spectra for positions at $\delta \leq -25°$ spaced 0°.5 in galactic latitude $b$ and approximately 0°.5/cos($b$) in galactic longitude. Each spectrum consists of 1008 values of antenna temperature for LSR velocities spaced 1.047 km s⁻¹ between -528 and 527 km s⁻¹. This velocity range was limited to ±459 km s⁻¹ after reduction. The observation cells were grouped in grids of 5 x 5 points. Four people performed the observations and data reduction, selecting the grids randomly according to the visibility with the IAR dish (hour angle range from -2 to +2 hours), and each person kept their own records of data reduction. Unfortunately those records were lost.

Each grid was usually observed completely during one turn, defined as a complete cycle including a) the observation of one of the 10 available calibration points at the beginning and at the end of the observation of the grid; b) the observation, at a central velocity of 1000 km s⁻¹(offset), of the first and last points of the grid, before and after the first and last point of the grid, respectively, and c) the observation of the points of the grid at a central velocity of 0 km s⁻¹. In addition, when visible, one of the cold sky positions (Sect. 6) was observed before the grid points.

The reduction consisted of:

1. The Fourier transform (FT) of the auto-correlator output.
2. The average of the offsets and the subtraction from the line profiles.
3. The fitting of a polynomial to the baseline and the subtraction from the profile.
4. The same (2,3) procedure for the nearest (in time) calibration profiles and determination of the calibration factors.
5. Multiplication of the line profile temperatures by the average of the selected calibration factors.

---

6. Correction of spikes and interferences in the line profile.

The survey that resulted from this procedure has not been made openly available because one of its aims was to complement the Leiden/Dwingeloo survey (LDS) on the northern sky (Hartmann, 1994; Hartmann & Burton, 1997) to produce a whole sky HI survey, a joint effort called the Leiden/Argentine/Bonn (LAB) survey. A description is given in the same volume of this journal (Kalberla et al., 2005). The observational parameters are quite similar but to make both surveys compatible it was necessary for both to be corrected for stray radiation. The LDS has stray radiation correction (SRC) included, but our survey does not and the effect of this difference is noticeable when merging the maps.

Analysis of the data, however was not simple because the reduced spectra of the southern survey could not be used for SRC since the baseline corrections performed on them might have affected features due to stray radiation.

It was necessary to recover the original profiles before any correction was done except the subtraction of the offset and the calibration. All the original spectra (immediately after the FT of the correlator output) were available, as well as many of the correlator outputs. There were, however, no records of the baselines and calibration factors used in the first reduction.

In the following section we describe the different processes applied to get the final data set.

## 2. Data reduction

### 2.1. The spectra for stray radiation correction

We subtracted the offset and applied the calibration factor to the 50980 original spectra. To do this, it was necessary:

– To identify, for each HI profile, all the spectra that were observed in the same observing turn.
– To select all the offset spectra suitable to be averaged (to decrease the noise), average them and subtract the average from the individual observed line profiles. This step produced the bandpass-corrected database.

The gain calibration and instrumental baseline was determined in the following way:

– A temporary polynomial was fitted to the bandpass-corrected spectrum using clean portions of the baseline and subtracted.
– The line profile area within a central narrow window was determined and compared to the area of the corresponding profile from the initial calibrated database.
– The bandpass-corrected spectrum was multiplied by this ratio and the gain factor was tabulated. If necessary the previous two steps were repeated, aiming to keep inconsistencies below 0.1%.

We compared the resulting spectra (corrected for bandpass, gain and instrumental baseline) with those from the database obtained by Arnal et al. (2000). The only acceptable differences were in the case of interference. These difference spectra were stored separately, as were the polynomial baselines and the recovered spectra (corrected for bandpass, gain and interference) *without* baseline correction.

From all the data sets created, only the latter two were needed for further reduction. The others were essential for the control and correction of the procedure in case of inconsistencies. In the next step the bandpass- and gain-corrected spectra were corrected for stray radiation.

### 2.2. Correction for stray radiation

Observations with radio telescopes result in antenna temperatures $T_a$ as a convolution of the true temperature distribution $T$ on the sky with the beam pattern $P$ of the antenna

$$T_a(x,y) = \int P(x - x', y - y')T(x', y')dx'dy'. \tag{1}$$

Equation 1 is a simplification. In general, it is time dependent, spherical coordinates should be used and the integration needs to be extended over the observable part of the sky and the ground where it is reflecting or radiating. In Eq. 1 the pattern $P$ of the antenna is normalized as

$$\int P(x,y)dxdy = 1 \tag{2}$$

and we conveniently may split the pattern into the main beam area (MB) and the stray pattern (SP)

$$T_a(x,y) = \int_{MB} P(x - x', y - y')T(x', y')dx'dy' + \int_{SP} P(x - x', y - y')T(x', y')dx'dy'. \tag{3}$$

Defining the main beam efficiency $\eta_{MB}$ of the telescope as

$$\eta_{MB} = \int_{MB} P(x,y)dxdy \tag{4}$$

we may rewrite Eq. 3 as

$$T_B(x,y) = \frac{T_a(x,y)}{\eta_{MB}} - \frac{1}{\eta_{MB}} \int_{SP} P(x-x', y-y')T(x',y')dx'dy'. \tag{5}$$

Here $T_B$ is the so-called brightness temperature. If we had an ideal telescope, with a main beam and no further sidelobes, we could observe $T_B$ as a convolution between the main beam of the telescope and the sky temperature $T$. For observations with a real telescope we can derive $T_B$ if we are able to calculate the contribution from the stray pattern (SP).

There are two problems in calculating the stray radiation: Eq. 5 requires knowledge of the "true" temperature $T$. This is unknown and the only alternative is to use $T_a$ instead. Also, the antenna pattern $P$ is usually not known in all its details. In this case we substitute the missing details by model assumptions.

Under the assumption that the antenna pattern $P$ can be described well enough, Bracewell (1956) proposed to derive $T_B$ by successive approximations. The first approximation would be to insert $T_a$ in Eq. 5. Kalberla (1978) has shown that it is possible to calculate the stray radiation in one step. It is possible to modify Eq. 5 by replacing $T$ with $T_a$ and $P$ by $Q$, the so-called resolving kernel function.

$$T_B(x,y) = \frac{T_a(x,y)}{\eta_{MB}} - \frac{1}{\eta_{MB}} \int_{SP} Q(x-x', y-y')T_a(x',y')dx'dy'. \tag{6}$$

Such a solution applies whenever the main beam efficiency of the telescope $\eta_{MB} > 0.5$. $Q$ can be calculated by successive approximations (Kalberla, 1978; Hartmann et al., 1996, Eq. 14). The Leiden/Dwingeloo survey (LDS) (Hartmann & Burton, 1997) was corrected in this way.

For the present survey we initially chose this solution. We first derived the resolving kernel function $Q$. The stray radiation was then calculated from Eq. 6, using antenna temperatures $T_a$ as observed and supplemented by antenna temperatures from the LDS. In a second iteration the derived brightness temperatures, as well as those from the LDS, were inserted in Eq. 5 to obtain the most accurate solution. Our calculations were supplemented by a similar correction of the LDS. For details we refer to Kalberla et al. (2005). In the following we use the term "LDS second edition" for this version of the LDS.

## 2.3. The antenna diagram

Details of the antenna diagram, as required to correct the observations for stray radiation, have been determined partly by modeling and partly by direct measurements. The response of the feed horn was measured in the lab of the MPIfR, Bonn, Germany. Taking in addition blocking by the feed support legs and the prime focus cabin into account, we used Fourier transform techniques (Bracewell, 1956) to calculate the aperture distribution and the corresponding far field pattern. Fig. 1 displays the antenna diagram according to this model within a radius of $14°\!\!.4$. This region was used to correct the stray radiation from the near sidelobes after re-binning and averaging the pattern within 468 cells in polar coordinates. Fig. 2 shows two cuts through the sidelobes along the major axes in RA (dashed) and DEC (solid lines).

The near sidelobe structure as displayed in Fig. 1 is rather complicated. This is because of the complicated geometry of the three-pod structure which deviates up to $5°\!\!.6$ from the ideal $120°$ symmetry. This causes asymmetries in the diagram. The main features of the diagram resemble that of the Dwingeloo telescope as measured by Hartsuijker et al. (1972). Only a minor part of the near sidelobe range could be verified observationally. At distances of 2 - $6°$ off the main beam the Sun was used. Sidelobes with levels up to -40 dB were found, roughly in agreement with the pattern from the model. The width of the main beam, FWHM = $30°\!\!.0$ in RA and FWHM = $29°\!\!.5$ in DEC, was measured with several point sources.

For the far sidelobe region (distance $\gtrsim 14°\!\!.4$ from the main beam) the positions of the main sidelobes were determined from the geometry of the telescope. Most important are the stray cones with a radius of $29°$ and the spillover region up to distances of $120°$ from the main beam. For a detailed discussion of the effects caused by these sidelobes we refer to Hartmann & Burton (1997). We verified the positions of most of these features with a transmitter, the mean levels were determined by fitting the observed time variability of the profiles similar to Kalberla, Mebold, & Reich (1980); Kalberla, Mebold & Velden (1980).

It was noted by Hartmann & Burton (1997, Fig. 24) that a minor part of the stray radiation must originate from reflection from the ground around the telescope. Such reflections have been determined by Kalberla et al. (1998) and have been removed from the second edition of the LDS. Angles of incidence and directions of specular reflection caused by the area around the 30-m telescope were estimated from land surveys and from photos taken from the prime focus of the telescope. The average albedo of the reflecting areas ($\sim 30\%$) was determined by minimizing the stray radiation. This value is rather crude. More accurate estimates would depend on weather as well as the height of the vegetation but such data have not been recorded.

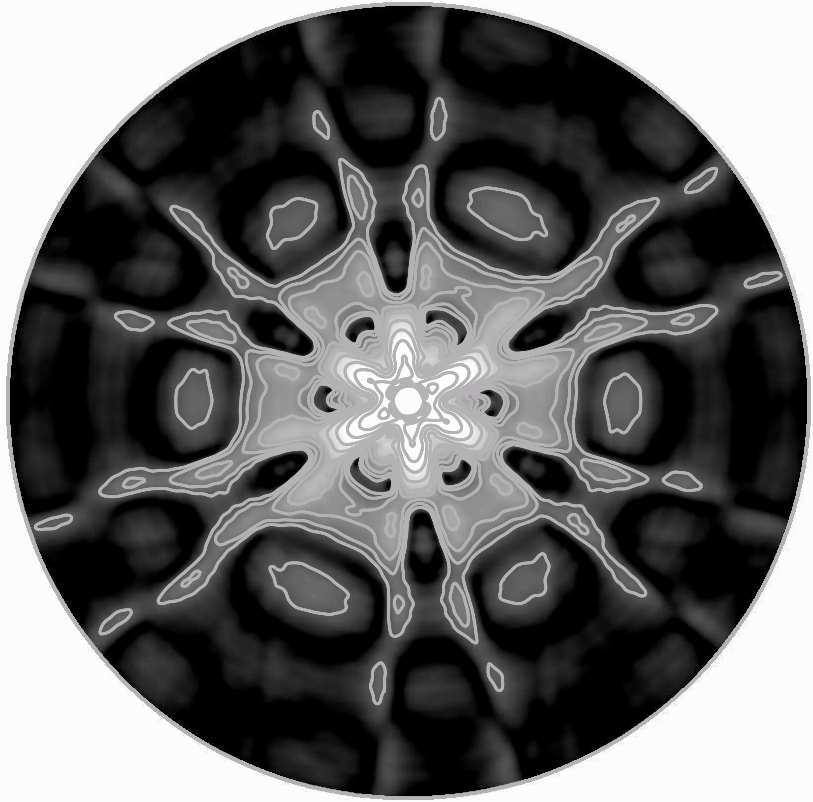

**Fig. 1.** Beam pattern of the 30-m telescope within a radius of 14°.4 as derived from the aperture distribution. The gray scale is between -60 and -30 dB, contours are plotted from -50 to -20 dB in steps of 3 dB.

## 2.4. Baseline correction

As discussed in Sect. 2.1, all profiles have been calibrated before eliminating the contribution from stray radiation. The correction for the instrumental baseline was the last step. The reason for this is that the stray radiation from the antenna diagram frequently causes extended profile wings (see Figs. 6 to 10 for examples). Such wings may easily be eliminated when correcting $T_a$ spectra for baseline defects. After removal of stray radiation such profiles would suffer from extended baseline regions below zero.

To avoid such spurious features it is important to keep any baseline corrections as the last reduction step. However, we used the baseline correction as applied to the observed antenna temperatures and described in Sect. 2.1 as a first guess. We then iteratively determined the baselines by fitting polynomials of order 1 to 4, each time searching for genuine HI features which in fitting should not be mistaken for baseline effects. After applying the $4^{th}$ order polynomial fit we frequently found residual profile wiggles suggesting standing wave problems. We removed these by a sine wave fit, allowing two sine waves at a time. Initial guesses for the sine wave fits were determined by calculating a mean standing wave for all observed profiles.

In the presence of standing waves a polynomial fit of $4^{th}$ order may misinterpret parts of a sine-wave as a polynomial. To avoid biases of this kind we repeat the final $4^{th}$ order polynomial fit, using the the same parameters as before, but this time after subtraction of the sine waves. To avoid any biases in fitting a baseline to weak profile wings the first zero transition on both sides of the main line was determined. This region, flagged as genuine emission, was extended on both sides by an additional 40 channels in the final fit.

The baseline procedure as described above was run automatically. The same baseline procedure but with minor modifications concerning the edge channels was used by Kalberla et al. (2005) for the second edition of the LDS. However, a minor fraction of the observed fields in the southern sky ($\lesssim 1 - 2\%$) had spurious residual baseline ripples. Attempts to repeat the bandpass calibration according to Sect. 2.1 were only partly successful. Therefore we decided to use additional baseline constraints after inspection of the sky surrounding these fields.

## 2.5. Calibration

Calibrating HI line observations aims for data with a temperature scale that is independent of the telescope used. The IAU recommended the standard fields S8 (primary), S7 and S9 (secondary) for an inter-comparison of the temperature scales between different telescopes (van Woerden, 1970). During observations, the survey was calibrated on a regular basis against the standard position S9. The telescope has an equatorial mount and a limited hour angle range ($\pm 2^h 0$). Since S9 is not observable all the time, ten additional tertiary calibrators were used to ensure a consistent temperature calibration under all conditions (Arnal et al., 2000, Table 2).

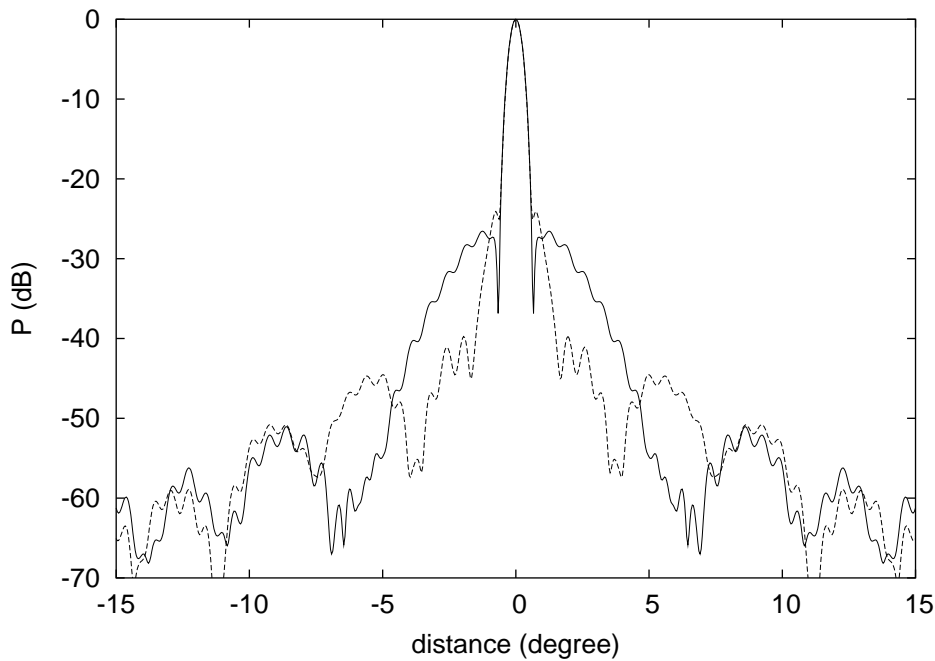

**Fig. 2.** Sidelobes for the 30-m telescope along the major axes of Fig. 1: RA (dashed) and DEC (solid lines)

Observed antenna temperatures are affected by the beam shape, therefore side-lobe effects need to be taken into account for calibration (Williams, 1973). After correcting the observations for stray radiation, we compared our data base with the second edition of the LDS (Kalberla et al., 2005). Based on 7256 common positions we compared column densities for $-200 < V < 200$ km s$^{-1}$, resulting in the following linear regression with a correlation coefficient of 0.997

$$N_H(IAR) = (1.021 \pm 0.003)N_H(LDS) + (3.2 \pm 3.7)10^{18}\text{cm}^{-2} \tag{7}$$

Initially (Arnal et al., 2000), the temperature scale of the IAR survey was matched to the scale proposed by Williams (1973). A scale error of 2%, found after correcting for stray radiation, is not surprising. It needs to be compared to the uncertainties of 7.5% due to the necessary beam deconvolution as estimated by Williams (1973). Fig. 5 displays the mean brightness temperature profile for S9 after correction for stray radiation. For comparison we include profiles for the total stray radiation (upper dashed line) and the stray radiation within the far sidelobe range (lower dashed line).

We match the final IAR brightness temperature scale to the LDS temperature scale (second edition) by applying a scale factor of 0.98 according to relation 7 for all $T_B$ data. In Table 1 we summarize our results for all calibrators used. This table may be compared to Table 2 of Arnal et al. (2000), valid for the direct calibration of the observations without a detailed correction for stray radiation. Due to stray radiation effects the ratios $N_H(T_B)/N_H(T_a)$ or $T_{max}(T_B)/T_{max}(T_a)$ show considerable fluctuations.

The LDS brightness temperature scale in the first and second edition of the survey (Hartmann et al., 1996; Kalberla et al., 2005) was matched to the Effelsberg brightness temperature scale (Kalberla, Mebold & Reif, 1982). This allows us to cross-check our calibration. Brüns (2003); Brüns et al. (2004), mapping the IAU calibrator field S8, have also matched the calibration of the Parkes telescope to the Effelsberg temperature scale. Convolving calibrated Parkes data around position S9 to the IAR beam shape, Brüns (private communication) derived a temperature scale which deviates from our scale by 0.1% only. The uncertainties for such a comparison are probably five times larger, the agreement is excellent and we conclude that the IAR brightness temperature scale is well established.

After matching the IAR temperature scale to the LDS (second edition) temperature scale we compare in Fig. 3 the column densities measured with both telescopes at common positions for $-400 < V < 400$ km s$^{-1}$. In addition, we allow in Fig. 4 an inter-comparison with column densities derived from the Bell Labs survey (Dickey & Lockman, 1990). In the latter case we use the widely distributed software to search for a column density in the Bell Labs database at a given position of the LDS. This introduces some scatter due to position uncertainties and mismatch in spatial resolution. Nevertheless, the scatter diagram shows not only statistical errors but also systematic offsets in the Bell Labs survey (up to 30%) due to sidelobe effects. For comparison, systematic inconsistencies in our database, as visible in Fig. 3, are approximately one order of magnitude less severe.

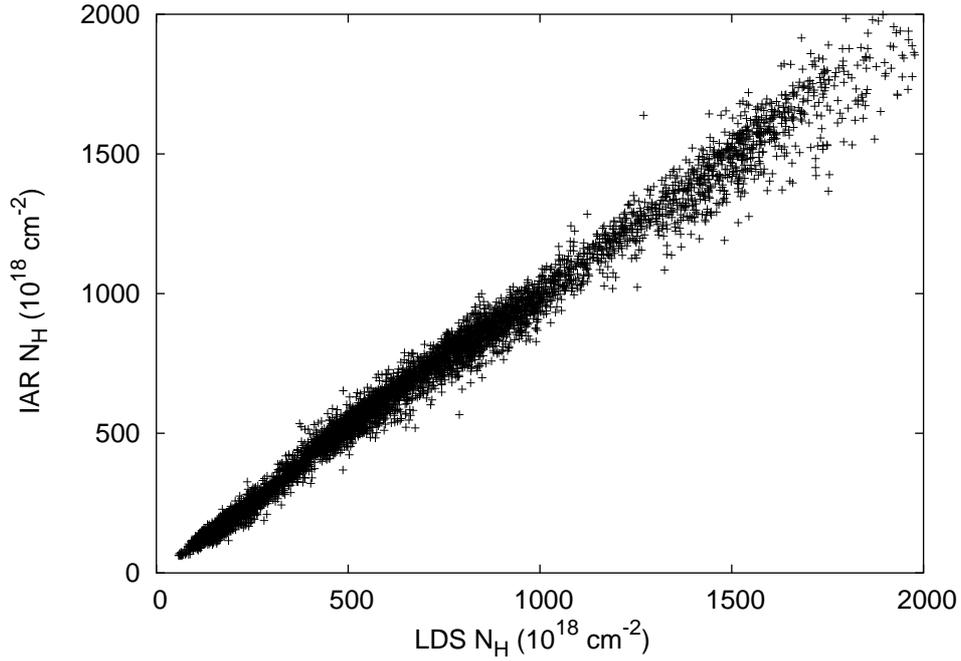

**Fig. 3.** Comparison of column densities as observed with the Dwingeloo telescope (second edition) and the 30-m telescope of the IAR at Villa Elisa at common positions for $-30° < \delta < -25°$.

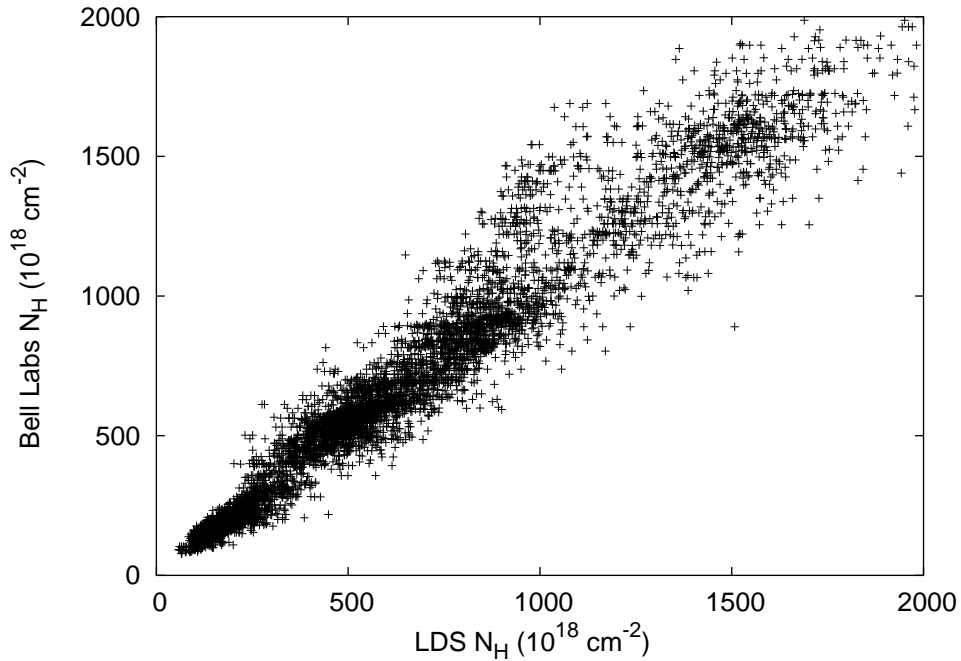

**Fig. 4.** Inter-comparison of column densities derived from the Bell Labs data base with LDS (second edition) column densities at the same positions as used in Fig 3.

## 2.6. Regions with low brightness temperatures

Observations at regions with low brightness temperatures are most critical since they are significantly contaminated by stray radiation. At such positions seasonal fluctuations are most pronounced, and the determination of the baseline is quite a problem since stray radiation may cause extended profile wings. In Figs. 6 to 10 we plot the mean corrected profiles together with the mean contributions from stray radiation, most of which is caused by the far sidelobes (lower dotted lines). The observational parameters for these regions have been included in Table 1.

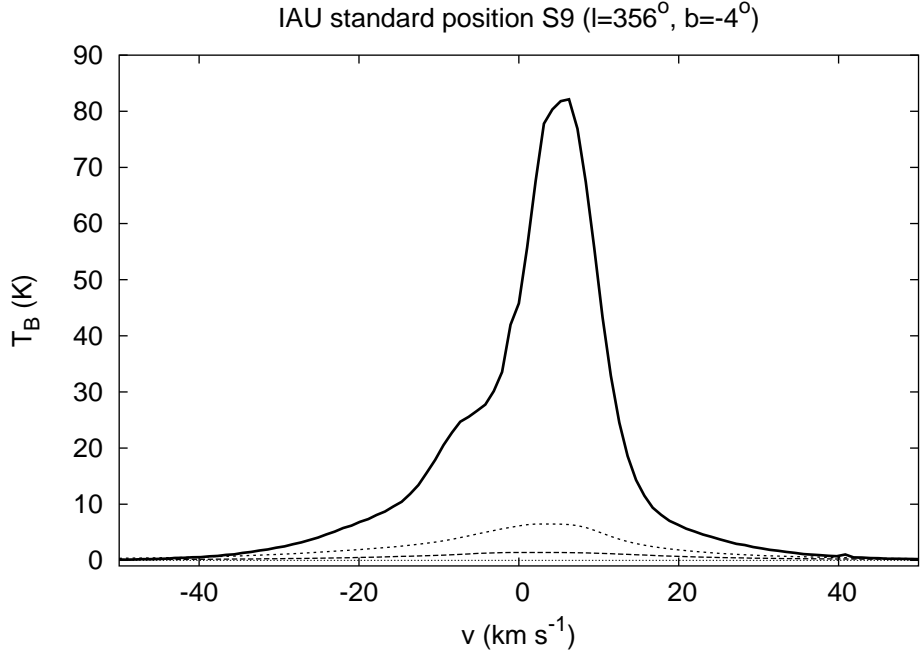

**Fig. 5.** Brightness temperature at the IAU standard position S9 (solid line). The profiles are included for the total stray radiation which was subtracted (upper dashed line) and the stray radiation contribution originating from the far sidelobes (lower dashed line).

**Table 1.** Calibration points and cold sky regions for the IAR survey

| Point | Gal. long. | Gal. lat. | Peak $T_b$ (K) | Prof. area (K km s$^{-1}$) | Vel. range (km s$^{-1}$) |
|---|---|---|---|---|---|
| IAR0(≡S9) | 356°.00 | −4°.00 | 82.5±1.1 | 905.5±2.9 | −1.5 to +15.2 |
| IAR1 | 4°.00 | −25°.00 | 46.2±1.4 | 278.0±2.1 | +0.5 to +12.0 |
| IAR2 | 304°.55 | −28°.87 | 41.3±0.9 | 271.9±3.2 | −3.6 to +7.8 |
| IAR3 | 304°.65 | −33°.23 | 47.0±1.1 | 282.7±2.4 | −2.6 to +8.9 |
| IAR4 | 298°.56 | −37°.75 | 41.8±1.3 | 262.4±2.9 | −3.6 to +7.8 |
| IAR5 | 301°.32 | −28°.73 | 33.4±0.9 | 261.4±3.9 | −4.7 to +6.8 |
| IAR6 | 219°.50 | −12°.00 | 82.3±2.5 | 649.8±2.0 | −1.5 to +9.9 |
| IAR7 | 247°.00 | +17°.00 | 66.2±1.8 | 490.0±4.3 | −8.9 to +2.6 |
| IAR8 | 290°.00 | +5°.00 | 67.0±1.6 | 673.5±7.6 | −16.2 to −4.7 |
| IAR9 | 306°.00 | +10°.00 | 60.1± 1.5 | 464.3±4.7 | −17.3 to −5.8 |
| IAR10 | 330°.50 | +10°.00 | 51.5±1.5 | 377.3±3.2 | −2.6 to +8.9 |
| Cold1 | 233°.23 | −27°.54 | 3.01±0.20 | 105.5±6.8 | −100. to +150. |
| Cold2 | 347°.31 | −75°.54 | 2.67±0.22 | 54.3±5.2 | −100. to +150. |
| Cold3 | 221°.98 | −52°.01 | 1.38±0.15 | 43.0±3.7 | −100. to +150. |
| Cold4 | 11°.44 | −59°.82 | 1.82±0.14 | 52.9±3.5 | −100. to +150. |
| Cold5 | 261°.00 | −40°.00 | 0.80±0.16 | 38.6±4.1 | −100. to +150. |

In total 268 profiles at the cold sky positions 1 to 5 have been analyzed. We find typical uncertainties of 0.15 to 0.2 K for the peak brightness temperatures. Such uncertainties may reflect some residual problems with stray radiation. Uncertainties in the instrumental baseline are more significant. As is obvious from Figs. 6 to 10 the baseline fit usually needs to interpolate at least a velocity range of ±50 km s$^{-1}$. It is quite problematic to distinguish possible weak profile wings from instrumental baseline uncertainties. We therefore forced the baseline fitting routine, described in more detail in Sect. 2.3, to disregard at least a 40 km s$^{-1}$ broad region on both sides of the main HI emission when fitting the baseline. Under such conditions it is quite acceptable that the necessary interpolation of the baseline leads to uncertainties which are enhanced by a factor of 2 to 3 over the typical rms uncertainties of 0.07 K as determined outside regions with line emission. For the determination of the total HI column densities we derive corresponding typical uncertainties of $10^{19}$ cm$^{-2}$.

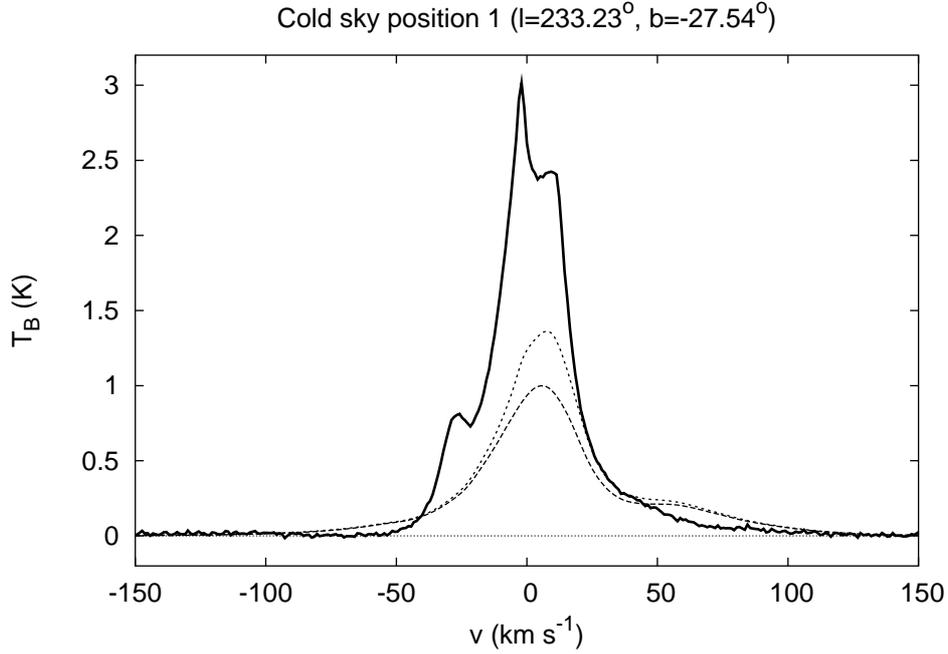

**Fig. 6.** Mean brightness temperature at the cold sky position 1 (solid line). Included are the mean profiles for the total stray radiation which was subtracted (upper dashed line) and the stray radiation contribution originating from the far sidelobes (lower dashed line). Note the extended profile wings due to stray radiation.

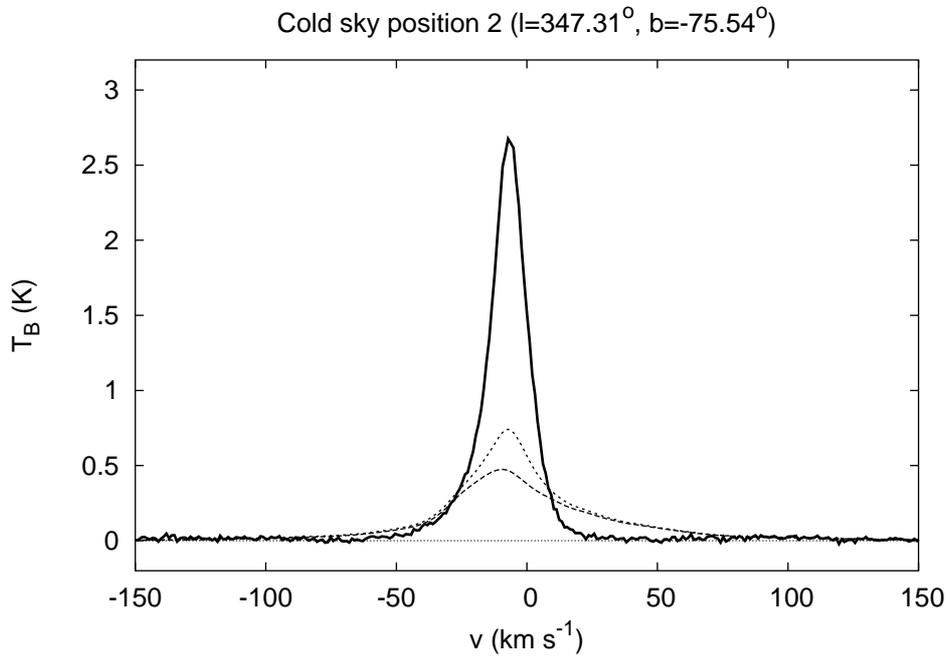

**Fig. 7.** Mean brightness temperature and stray radiation at the cold sky position 2.

## 2.7. Limitations

Using a rather restrictive calibration scheme, removing instrumental stray radiation, reflections from the ground and baseline defects, we have tried to minimize systematic errors in our survey. There are still some residual problems. The most important are uncertainties due to interference. Many profiles have been re-observed for this reason, superseding the Arnal et al. (2000) database. Weaker spikes have been removed. 1-2% of the final survey data still may contain some errors, such as interference-induced profile components, scale errors and residual baseline defects. Most of these problems are easily detectable as patchy features in the channel maps. Most of the profiles, however, should have scale errors below 1 - 2% and the baseline should be accurate to 20 - 30 mK, resulting in errors for the total column density of $10^{19}$ cm$^{-2}$.

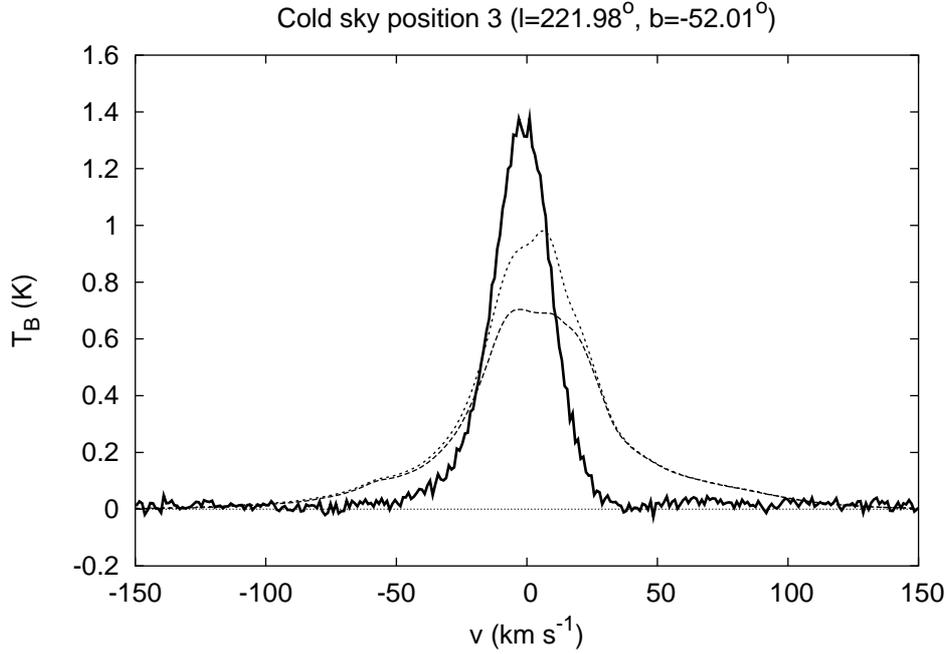

**Fig. 8.** Mean brightness temperature and stray radiation at the cold sky position 3.

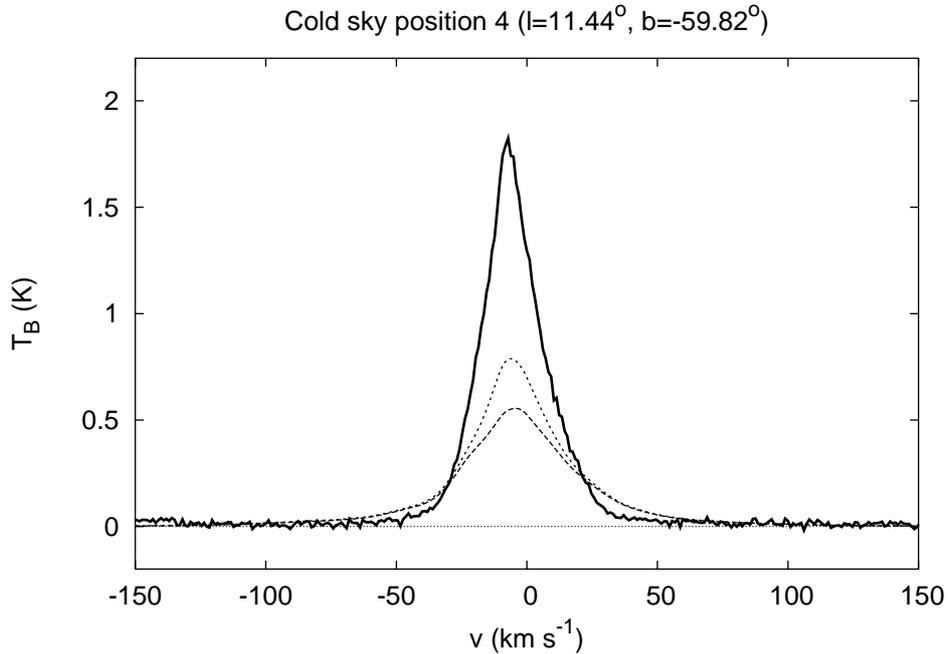

**Fig. 9.** Mean brightness temperature and stray radiation at the cold sky position 4.

After completing the final data reduction we detected an additional problem, a jitter of the observed center velocities. We found statistical uncertainties of 0.3 km s$^{-1}$ for the second moments of our calibration profiles. This problem exists for all of our data and we were unable to detect any systematic trends. The most probable explanation for this problem is a general instability of the local oscillator system.

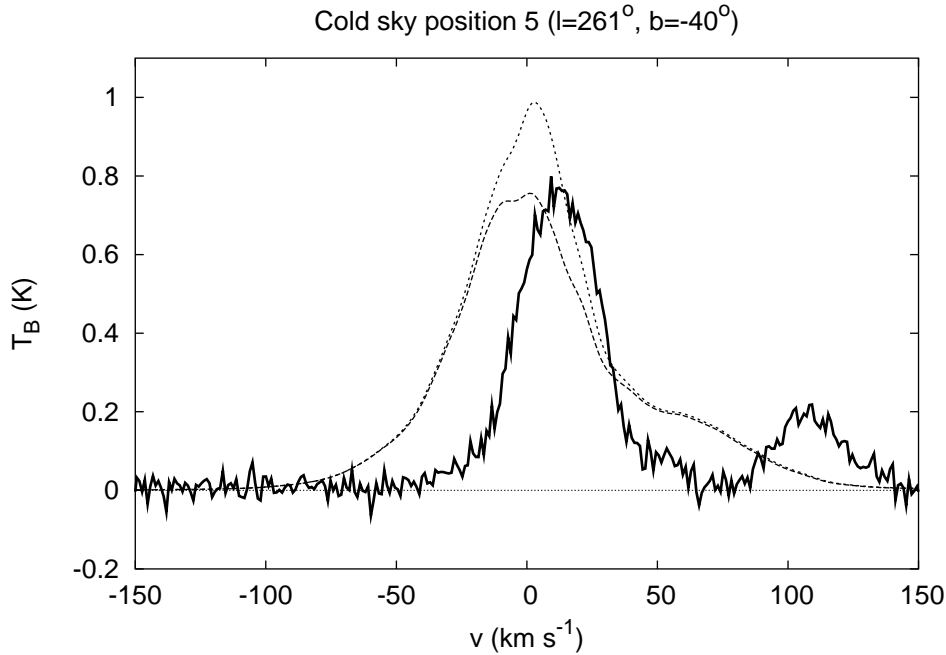

**Fig. 10.** Mean brightness temperature and stray radiation at the cold sky position 5.

## 3. Data

All 50980 observed profiles have been combined to a single FITS data cube. This cube covers $\pm 459$ km s$^{-1}$ with the original velocity resolution. For a regridding in spatial coordinates, a Gaussian function with a FWHM of $0.°3$ has been used. In this way the spatial resolution is degraded to $\sim 0.°6$ ($0.°58$ in RA and $0.°576$ in DEC but note that the sampling violates the Nyquist rate). The data are available at CDS. (********** give link to catalog VIII/75 here ************).

*Acknowledgements.* Part of this work was supported by the Consejo Nacional de Investigaciones Cinetíficas y Técnicas (CONICET) of Argentina under Project PIP 2277/00, also by the *Deutsche Forschungsgemeinschaft, DFG* project number Ka 1265/2–1. We are grateful to Ulrich Mebold for continuous support. We thank Christian Brüns for making Parkes calibration data available. Urmas Haud is acknowledged for running his Gaussian analysis on the survey database for quality control.